\documentclass[12pt]{iopart}

\usepackage{iopams}

\usepackage{graphicx}

\begin{document}
\title[Influence of photoexcitation depth on
luminescence spectra of bulk GaAs\dots]{Influence of photoexcitation depth on
luminescence spectra of bulk GaAs single crystals:
application to defect structure characterization}

\author{V~A~Yuryev\footnote{E-mail: VYURYEV@KAPELLA.GPI.RU.},
V~P~Kalinushkin}
\address{General Physics Institute of the Russian Academy of Sciences,
38, Vavilov Street, Moscow, GSP--1, 117942, Russia}
\author{A~V~Zayats\footnote{E-mail: AZAYATS@UALG.PT. Present address: University
of Algrave, P--8000, Faro, Portugal.},
Yu~A~Repeyev\ and V~G~Fedoseyev}

\address{Institute of Spectroscopy of the Russian Academy of Sciences, Troitsk, Moscow region, 142092, Russia}



\begin{abstract}
The results of investigation
of bulk GaAs photoluminescence are presented
taken from near-surface layers of different
thicknesses using for excitation the light with the wavelengths which are
close but some greater than the excitonic absorption
resonances (so-called ``bulk'' photoexcitation). Only the excitonic and
band-edge luminescence is seen under the interband excitation,
while under the ``bulk'' excitation, the spectra are much more
informative. The interband excited spectra of all the
samples investigated in the present work are practically
identical, whereas the bulk excited PL spectra are different for
different samples  and  excitation depths and provide
the information on the deep-level point
defect composition of the bulk materials.
\end{abstract}

\section{Surface and quasi-bulk photoexcitation: two approaches to
luminescence characterization of semiconductors}
   Photoluminescence  (PL) technique is a powerful tool for the
investigation of defects
in semiconducting materials. In most cases,
the luminescence of semiconductors is investigated under the
interband excitation, so that electrons are excited from the
valence band to the conduction band. After the excitation, the
electrons and holes thermalize to the thermoequilibrium
distribution in the conduction and valence bands due to
electron--phonon and electron--electron (hole--hole)
interactions in a time less than 0.1\,ps for the holes in the
valence band and of about 1\,ps for the electrons in the
conduction band \cite{therm,nin}, and then recombine. Main
channels of the radiative recombination in semiconductors are
the following: free and bound exciton luminescence, band-edge PL
involving shallow acceptor-like defects, and luminescence via the
states of the deep-level defects or the defect associations.
All three types of the PL yield the information on the defect
structure of semiconductors.

\subsection{Do PL measurements show real deep-level defect structure
at band-to-band excitation?}
  In case of the interband excitation $(\hbar \omega > E_g)$,
the exciting light is absorbed in the region less than
0.1\,$\mu$m close to the surface. For this reason, the spectra of
the observed luminescence reflect mainly the defect structure of the
near-surface layer but not of the sample bulk, whereas the
information about the defect structure of the crystal bulk is of
primary interest for solving many fundamental and applied
issues. The case is that the procedures of preparation of
experimental samples or technological wafers\,---\,such as cutting,
abrading, and lapping\,---\,enter a great amount of defects in the
near-surface layer, and despite the following etching removes this damaged
layer, a considerable concentrations of non-uniformly
distributed defects entered by these procedures remains in the
near-surface layers\ \cite{mst}. Moreover, even if the
damaged layer is etched off up to the depth where the influence
of the former treatments seems to be practically absent,
the etching itself changes the defect structure of the
near-surface layer\ \cite{apl}, e.g. by removing the impurities,
intrinsic defects, precipitates and their agglomerations, which
are present in the crystal bulk, or by entering some new
defects. Besides, some sample treatments, such as annealing (or
those involving annealing), {\em etc}, lead to creation and/or
redistribution of the defects between the crystal bulk and its
near-surface layer, so the defect structure of the near-surface
region becomes not identical to the bulk one and nonuniform.
{\em Therefore,
interband excited PL reflects neither the real defect structure
of the crystal bulk nor its changes after the exposures given to
the samples.}

   Unfortunately, in the PL measurements and particularly in the
PL mapping of semiconducting materials neither the above
considerations nor the circumstance, that the near-surface layer
modified with special treatments\,---\,e.g. with etching\,---\,is
investigated rather than the material bulk, are often taken into
account. (It should be noted that we do not mean etching
simulating one of the technological steps of a device
fabrication~---~in this case the procedure used would be correct,
but only etching aimed to reduce the non-radiative
recombination beneath the surface and enlarge the PL
efficiency\,\footnote{Note that this remark is valid for some
other methods
of investigations of semiconductors such as EBIC, OBIC, DLTS,
{\em etc}, as well.}.)
The inferences valid for the particular way of
surface preparation are often spread to whole the material
bulk or near-surface regions subjected to other kinds of
pre-experimental treaments.

\subsection{Photoluminescence at band-tail excitation: a way to
obtain true information on subsurface defect structure}
Nonetheless some modifications of the routine PL measurements
might easily be done, which enable obtaining more correct
information about the defects in the bulk of crystals.  To study
the photoluminescence of the crystal bulk, the light with a
wavelength close but some greater than the excitonic absorption
resonances for the PL excitation can be used. At these
wavelengths, the absorption is expected to be still effective
for PL excitation due to the band tails. At the same time, the
absorption coefficient for this light is not so large as for the
interband excitation hence the absorption length is large enough
to excite a bulk of a sample (a layer with effective
thickness up to 100\,$\mu$m might easily be studied).

All three types of the luminescence mentioned above for the
interband photoexcitation can be excited under the ``bulk''
excitation. Nevertheless, the excitonic and band-edge PL
generated in the sample bulk could hardly be seen in the PL
spectra due to absorption and re-radiation by the near-surface
layer because of the large absorption coefficient at these
wavelengths. It means that the surface generated excitonic and
band-edge PL is always seen in the spectra irrespectively to the
``surface'' or ``bulk'' excitation is used.  Another situation takes place
for the deep-level luminescence. The wavelengths of this
luminescence are far from the band gap and the absorption is weak
at these wavelengths, so the spectrum of the bulk-excited
deep-center luminescence is less affected by the surface layer.
It gives one the possibility to investigate the defect structure
in a bulk of samples by means of the deep-level-center PL
excited by the bulk-absorbed light.

In the present paper, the authors strive to demonstrate how
the band-tail excited PL might be used for the investigation of
the effect of sample processing on the deep-level structure of the
crystal bulk, and the vapour phase epitaxy (VPE) of GaAs as
well as the VPE-simulating annealing are taken as an example.

\section{Experimental details and basic result}
\subsection{PL setup}
The photoluminescence spectra were recorded using 20-ps
pulses from a tunable parametric oscillator consisting of two
DKDP crystals pumped by the second harmonic of the
YAG:Nd$^{3+}$--laser radiation\ \cite{lum}. The output radiation
wavelength could be continuously tuned from 370 to 1890\,nm.
The emission with the wavelengths of 580, 810, 835 and 845\,nm
was used for PL excitation.
The maximum energy of the pulses was of 0.1 and 0.5\,mJ at 580-nm
and 810-nm excitation wavelengths, respectively, and 1\,mJ at
835-nm and 845-nm excitation wavelengths. The unfocused beam was used
to avoid too high excitation densities at which nonlinear
effects could predominate. Relaxation processes in GaAs single
crystals were expected to be fast enough in order to consider
20\,ps-pulse excitation as quasistationary one\ \cite{therm}.
Taking into account the difference
between an absorbance at the excitation wavelengths and between
the energies of pump pulses, the concentration of the excited
electrons was estimated to be ranged from $10^{15}$ to
$10^{17}$\,cm$^{-3}$ under all the excitation wavelengths.
In average, the effective thicknesses of the
layers excited in our experiments were estimated to be of around
0.1, 0.5, 1, and 10\,$\mu$m at 580, 810, 835, and 845-nm
excitation wavelengths, respectively.

The spectra were taken at the temperature of 80\,K using
a liquid nitrogen cooled cryostat.
For the spectral analysis, a MDR--3 diffraction monochromator
(LOMO) and photomultiplier were used. The PL signal was averaged over 20
pump pulses whose intensity was in the preset range.

\subsection{Samples}
   To investigate the effect of the vapour phase epitaxy (VPE)
on the defect structure of the bulk substrate, undoped GaAs
epitaxial layers were grown using the trichloride VPE process
on substrates of chromium-doped LEC gallium arsenide.
\begin{figure}[th]
\begin{center}
\includegraphics[scale=3]{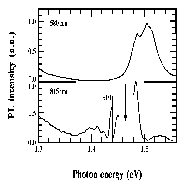}
\includegraphics[scale=3]{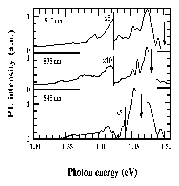}
\end{center}
\caption{Photoluminescence spectra of the as-grown GaAs:Cr wafer measured
at ``surface'' ($\lambda_{ex} = 580\,$nm) and ``bulk''
($\lambda_{ex} = 845\,$nm) excitations.}\label{f1}
\caption{PL spectra of the initial GaAs wafer obtained at different
excitation depths (excitation wavelengths are indicated in the
plot).}\label{f2}
\end{figure}
The wafers intended for epitaxial growth were cut from the
same ingot.
The dislocation density in the substrates was $10^4$--$10^5$\,cm$^{-2}$
and their resistivity was in excess of $10^8\,\Omega \,$cm.  The
substrate surfaces were oriented at angles of 2\,--\,6$^{\circ}$ with
respect to the (100) plane. The thicknesses of the substrates
were 300\,$\mu$m. The thicknesses of the films grown were about 6\,$\mu$m,
the film growing rate was 0.1\,$\mu$m/min, the growth temperature
was 720$^{\circ}$C. The carrier concentration in the epilayers did
not exceed $10^{14}$\,cm$^{-3}$ at room temperature.

   Some of the samples were subjected to a thermal treatment alone at
720$^{\circ}$C in a hydrogen atmosphere, which simulates the VPE
procedure.  The annealing lasted 1\,h. Reference samples were cut
from the same ingot being usually adjacent wafers in the ingot to the
processed ones.

   The samples were etched in the liquid etchant before the PL
measurements, so as the epilayers in the samples subjected to
VPE as well as about 10\,$\mu$m thick near-surface layers in the
samples, which were not subjected to the VPE process, were
etched off.

   The defects in the analogous samples after the same exposures
were previously investigated by means of the low-angle mid-IR-light
scattering technique (LALS)\ \cite{mse}.

\subsection{Basic result}
   As it is clearly seen from Fig.\,\ref{f1} where the spectra for the
\begin{figure}[t]
\begin{center}
\includegraphics[scale=3]{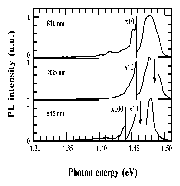}
\includegraphics[scale=3]{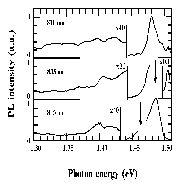}
\end{center}
\caption{PL spectra of the GaAs wafer after VPE obtained at different
excitation depths (excitation wavelengths are indicated in the
plot).}\label{f3}
\caption{PL spectra of the annealed GaAs wafer obtained at different
excitation depths (excitation wavelengths are indicated in the
plot).}\label{f4}
\end{figure}
as-grown GaAs sample are presented under the interband (580\,nm)
and band-tail (845\,nm) excitations,  only the excitonic and
band-edge luminescence is observed under the interband excitation,
while under the bulk excitation the spectrum is much more
informative.  The interband excited spectra of all the
samples investigated are practically
identical, whereas the influence of the sample treatment is
clearly seen under the band-tail excitation (Figs.\,\ref{f2}--\ref{f4}).
Three groups of bands are observed in the
spectra of the band-tail excited PL: excitonic and hot PL
($\hbar \omega\, ^{_>}_{^{\sim}}\, 1.49$\,eV),
band edge luminescence ($1.46\, ^{_<}_{^{\sim}}\, \hbar \omega\, ^{_<}_{^{\sim}}\,
1.49$\,eV), and deep centers related luminescence ($\hbar \omega\, ^{_<}_{^{\sim}}
1.46$\,eV).  The relative intensities and spectral positions of the PL bands
depend on the excitation wavelength as well as the wafer treatment.
Excitonic-related and hot photoluminescence depends weakly on the excitation
wavelength, that reflects its surface nature, but this PL is very sensitive to
the sample treatment.

{\em This is a basic result of the current work from which a technique
for correct analysis of the subsurface layer deep-level defect
structure can be easily derived in application to bulk semiconductors
and thick films.}

The intensity of
the deep-level related PL is not always a straightforward measure of
the center concentration. The ratio of the extrinsic
to intrinsic luminescence components is to be used to compare the
center concentrations from the luminescence intensity\ \cite{dean}.
For this reason, all the presented spectra are normalized on the
band edge luminescence intensity. The excitation
intensities varied from sample to sample and from one
excitation wavelength to another by less than 8\% during the
measurements, therefore the difference between the monomolecular
and bimolecular types of the deep-level and interband types of
recombination is expected to be small enough.  So we can use the
normalized PL intensities from the spectra of different samples
to compare the defect concentrations in the samples.

   Not only the intensities of the corresponding luminescence bands but
also the PL-bands wavelengths reflect the changes of the composition
and/or concentration of the defects in the bulk and in the
near-surface region because of the formation of different
complexes and defect associations in the bulk as well as close
to the surface that leads to the shift and/or splitting of the PL
bands.  Annealing or etching of the wafers can result in the
changes of the intensities and the spectral positions of the PL
bands because the treatments change the concentrations and
compositions of defects as well as the defect associations.

\section{Example of defect structure characterization: defect redistribution
as a result of vapor phase epitaxy and simulating annealing of GaAs}
After the treatments, dramatic changes of the PL intensities were observed.
Relative intensity of the excitonic PL increased
compared to the band-edge PL, while the deep-level
related PL decreased, and the degree of this decrease
depended on the treatment.  The defect-related PL from the crystal bulk
decreased by about 6 times after the annealing, and more
than 10 times after the epitaxy (Figs.\,\ref{f2}\,--\,\ref{f4}), i.e.
during the treatments either the defect diffused to the surface,
either the compensation of the sample  changed which
led to the radiative recombination centers became
nonradiative, or the defects gathered in the precipitates,
where they could not participate in the radiative recombination.
The epitaxial growth resulted in the
decrease of the defect-related luminescence. This
decrease was stronger for PL at 845-nm excitation (about 100 times)
and weaker for that at 810-nm excitation (about 10 times).  The
810-nm excited PL from the annealed wafer and the epitaxy subjected one had
comparable intensities for the bands at 1.39 and 1.41~eV, while the
concentration of more deep centres close to the surface was much
lower after the epitaxy than after the annealing.

   Deep centers related PL had a strong dependence on the
excitation wavelength that reflected the nonuniform distribution
of the defects between the bulk and the near-surface layer
rather than the resonance excitation of the luminescence, since
all the excitation wavelengths were far from the locations of
direct deep-level-to-band transitions and this PL was excited via
the interband or band-tail absorption. When the excitation
changed from ``surface'' one to ``bulk'' one, the increase of the deep
centres related luminescence intensity with respect to the
band-edge PL intensity could be clearly seen for all the
investigated samples (Figs.\,\ref{f1}\,--\,\ref{f4}).

\section{Summary}
Photoluminescence at band-tail excitation is a fruitful way to obtain a true
information on a real subsurface deep-level defect structure of
bulk semiconductors and thick layers. It enables the investigation
of the defect distribution deep inside the crystal bulk as well as the
redistribution of defects in subsurface layers resulting from
sample treatments.\\

\end{document}